\documentclass[final,5p,times]{elsarticle}
\usepackage{xr}
\usepackage{amsmath}
\usepackage[hyphens]{url}
\usepackage{amssymb}
\graphicspath{ {./figures/} }

\makeatletter
\newcommand*{\addFileDependency}[1]{% argument=file name and extension
  \typeout{(#1)}
  \@addtofilelist{#1}
  \IfFileExists{#1}{}{\typeout{No file #1.}}
}
\makeatother

\newcommand*{\myexternaldocument}[1]{%
    \externaldocument{#1}%
    \addFileDependency{#1.tex}%
    \addFileDependency{#1.aux}%
}

\myexternaldocument{SM}

\journal{Chaos, solitons and fractals}

\begin{document}

\begin{frontmatter}



\title{Explosive dismantling of two-dimensional random lattices under betweenness centrality attacks}


\author[famaf, ifeg]{Nahuel Almeira}
\author[famaf, ifeg]{Juan Ignacio Perotti}
\author[ifeg]{Andrés Chacoma}
\author[famaf, ifeg]{Orlando Vito Billoni}

\affiliation[famaf]{
    organization={Facultad de Matem\'atica, Astronom\'{\i}a, F\'{\i}sica y Computaci\'on, Universidad Nacional de C\'ordoba},
    addressline={Ciudad Universitaria}, 
    city={Córdoba},
    postcode={5000}, 
    state={Córdoba},
    country={Argentina}
}
    
\affiliation[ifeg]{
    organization={Instituto de F\'{\i}sica Enrique Gaviola (IFEG-CONICET)},
    addressline={Ciudad Universitaria}, 
    city={Córdoba},
    postcode={5000}, 
    state={Córdoba},
    country={Argentina}
}

\begin{abstract}
In the present paper, we study the robustness of two-dimensional random lattices (Delaunay triangulations) under attacks based on betweenness centrality. Together with the standard definition of this centrality measure, we employ a range-limited approximation known as $\ell$-betweenness, where paths having more than $\ell$ steps are ignored. For finite $\ell$, the attacks produce continuous percolation transitions that belong to the universality class of random percolation. On the other hand, the attack under the full range betweenness induces a discontinuous transition that, in the thermodynamic limit, occurs after removing a sub-extensive amount of nodes. This behavior is recovered for $\ell$-betweenness if the cutoff is allowed to scale with the linear length of the network faster than $\ell\sim L^{0.91}$. Our results suggest that betweenness centrality encodes information on network robustness at all scales, and thus cannot be approximated using finite-ranged calculations without losing attack efficiency.
\end{abstract}

\begin{highlights}

\item Percolation transition on Delaunay graphs based on betweenness and range-limited betweenness centrality attacks. 
\item Finite-size scaling analysis and critical exponent derivation.
\item Continuous and discontinuous transitions depending  on the range of interactions.
\end{highlights}

\begin{keyword}
Complex networks \sep Explosive percolation \sep Betweenness 
\end{keyword}

\end{frontmatter}


%\begin{linenumbers}

%%%%%%%%%%%%%%%%%%%%%%%%%%%%%%%%%%%%%%%%%%%%%%%%%%%%%%%%%%%%%%%%%%%%%%%%%
%%%%%%%%%%%%%%%%%%%%%%%%%%%%%%%%%%%%%%%%%%%%%%%%%%%%%%%%%%%%%%%%%%%%%%%%%
%%%%%%%%%%%%%%%%%%%%%%%%%%%%%%%%%%%%%%%%%%%%%%%%%%%%%%%%%%%%%%%%%%%%%%%%%
\section{Introduction}

%The organization of many complex systems is restricted by spatial constrains. Power grids, transportation and mobility networks, the Internet and the human brain are some examples. Spatial networks, where nodes and edges are embedded in two or three dimensional space, become a natural model for studying such systems~\cite{Barthelemy2011}. 

The organization of many complex systems is restricted by spatial constraints. Power grids, transportation and mobility networks, the Internet, and the human brain are all examples of systems whose structure and evolution are influenced by geometrical aspects. Spatial networks, where nodes and edges are embedded in two- or three-dimensional space, become a natural model for studying such systems~\cite{Barthelemy2011}. 

Realistic modeling of embedded networks is often attained by employing different random spatial networks. In the physics literature, one of the most extensively studied models is the Voronoi tessellation and its dual, the Delaunay triangulation (DT), also known as random lattice~\cite{Becker2009,Kirkley2018}.
%
%For building realistic embedded networks, tessellations and triangulations are often employed, as both, the Voronoi tessellation and its dual---the Delaunay triangulation---have been extensively used in the description of complex systems~\cite{Becker2009,Kirkley2018}.
It is worth mentioning that the distribution of local measures such as node degree, transitivity, and assortativity, which are widely used in complex network characterizations, become less informative in the case of spatial networks. For example, planar spatial networks usually exhibit a centered degree distribution,
and in a triangulation, transitivity is trivially maximized. Instead, other measures such as topological and geometrical distances become more relevant for these types of networks. In particular, a contrast between local and global measures can be expected in these kind of networks.    

Among the multiple aspects that can be addressed in the study of spatial complex networks, robustness is a topic that has drawn much attention for its theoretical and practical implications. In general, robustness refers to the ability of a system to maintain its functions when one or more of its parts are compromised. In the context of networks, connectivity plays an important role in terms of robustness as, for instance, the breakdown of structural connectivity is frequently followed by a systemic failure~\cite{Buldyrev2010}.

There is a vast literature where the robustness of DT, and other spatial networks,
is addressed~\cite{Sykes1964,Bollobas2006,Becker2009,Melchert2013,Norrenbrock2016,Norrenbrock2016a,DeOliveira2008, Ding2014NumericalLattices}  within the framework of percolation theory~\cite{StaufferBook}. 
The percolation thresholds of both site and bond percolation on two-dimensional DT are very well known~\cite{Sykes1964,Bollobas2006,Becker2009}, and there is evidence supporting that both processes belong
to the same universality class of random percolation on two-dimensional regular lattices~\cite{Hsu1999PercolationDuals,McCarthy1987InvasionLattice}.
In addition to random percolation, DT networks have also been studied in the context of centrality-based attacks. The idea is to assess how robust the network is when nodes considered important, or central, are deliberately removed. For instance, two different attack strategies are analyzed by Norrenbrock et al.~\cite{Norrenbrock2016}, each based on a different centrality measure.
In one of them, nodes with the highest degree are sequentially removed. Here, the network breaks faster than removing nodes randomly, but the nature of the transition does not change, as it belongs to the random percolation universality class. In the other attack strategy, based on node betweenness~\cite{Freeman1977}, the network breaks even sooner, and the percolation threshold is shown to be equal to zero in the thermodynamic limit. Nevertheless, the authors cannot determine the universality class of this transition.

In this work, we present a detailed study of the percolation transition induced by attacks based on initial node betweenness centrality.
We show that the transition occurs when a sub-extensive fraction of nodes is removed, as in the sequential version of the attack.
Moreover, the nature of the transition is compatible with a first-order transition, which allows us to frame the process in terms of explosive percolation~\cite{DSouza2019}.
In addition to this attack strategy, we employ a series of attacks based on approximate values of betweenness, where only limited-range paths are considered. In these cases, transitions are continuous and occur at non-zero values for the control parameter.

The paper is organized as follows. In Section \ref{sec:Methods} we describe the Delaunay triangulation and the attack strategies employed, and introduce the theoretical tools that we use to analyze the percolation transitions.
In Section \ref{sec:Results} we present our main results, which are then contextualized and further discussed in \ref{sec:Discussion}. Finally, we detail our conclusions in \ref{sec:Conclusions}.

%%%%%%%%%%%%%%%%%%%%%%%%%%%%%%%%%%%%%%%%%%%%%%%%%%%%%%%%%%%%%%%%%%%%%%%%%
%%%%%%%%%%%%%%%%%%%%%%%%%%%%%%%%%%%%%%%%%%%%%%%%%%%%%%%%%%%%%%%%%%%%%%%%%
%%%%%%%%%%%%%%%%%%%%%%%%%%%%%%%%%%%%%%%%%%%%%%%%%%%%%%%%%%%%%%%%%%%%%%%%%
\section{Methods} \label{sec:Methods}

%%%%%%%%%%%%%%%%%%%%%%%%%%%%%%%%%%%%%%%%%%%%%%%%%%%%%%%%%%%%%%%%%%%%%%%%%
%%%%%%%%%%%%%%%%%%%%%%%%%%%%%%%%%%%%%%%%%%%%%%%%%%%%%%%%%%%%%%%%%%%%%%%%%
\subsection{Delaunay triangulation}\label{subsec:Delaunay}

The Delaunay triangulation is one of the most studied models of spatially embedded systems. Applications of this graph are known in diverse fields, such as the construction of ad-hoc wireless networks~\cite{Meguerdichian2001CoverageNetworks}, the modeling of cities~\cite{Kartun-Giles2019ShapeNetworks} and the study of phase transitions under quenched disorder~\cite{Barghathi2014PhaseDisorder}.  A Delaunay triangulation is constructed as follows; given a set of points $V$ in a d-dimensional space, a link between two nodes $u,\;v\in V$ is present if and only if there exists a d-dimensional sphere which embeds $u$ and $v$ but no other points. 
The DT is an example of an \emph{excluded region graph}, where connectivity is based on the absence of points in a region between two nodes. Other members of this class of spatial networks are the Gabriel graph, the relative neighborhood graph, and the euclidean minimum spanning tree, all of them subgraphs of DT. From a thermodynamic viewpoint, all these models have been shown to behave similarly. In particular, continuous percolation-related models defined on them share the same critical exponents~\cite{Kartun-Giles2019ShapeNetworks,Norrenbrock2016}.
We have chosen to work with DT because of its widespread applications, but expect our results to generalize to other kinds of spatial networks.

%%%%%%%%%%%%%%%%%%%%%%%%%%%%%%%%%%%%%%%%%%%%%%%%%%%%%%%%%%%%%%%%%%%%%%%%%
%%%%%%%%%%%%%%%%%%%%%%%%%%%%%%%%%%%%%%%%%%%%%%%%%%%%%%%%%%%%%%%%%%%%%%%%%
\subsection{Betweenness centrality}

Node centrality has been largely studied in the complex network literature and presents many interesting aspects. To begin with, the concept of centrality itself is not universally defined, as it depends on particular characteristics of the networks studied, the phenomena taking place on top of them, and the questions of interest.
In some cases, as in social networks, nodes that are more connected with their neighbors are considered the most influential ones. For these systems, metrics such as degree or collective influence~\cite{Morone2015} may represent good estimations of node centrality.
On the other hand, in technological networks such as road networks or the Internet, where traffic or information flows are present, central nodes are generally located in high traffic paths. Here, metrics as closeness~\cite{Sabidussi1966TheGraph} and betweenness~\cite{Freeman1977} are usually preferred. In the past years, a vast set of centrality measures have been introduced. Although each one describes different aspects of the nodes, and the networks in general, some of them are closely related and usually display a high correlation between them~\cite{Oldham2019}.

Betweenness centrality (BC) was independently proposed by~\cite{Anthonisse1971TheGraph} and~\cite{Freeman1977} in the context of social networks and its use has spread through the complex networks literature, with applications such as community detection~\cite{Newman2004}, network robustness~\cite{Holme2002}, and organization of cities~\cite{Kirkley2018}. It can be defined in the following way.
Let $\sigma(s, t)$ be the number of shortest paths connecting nodes $s$ and $t$ and let $\sigma_i(s, t)$ be the number of such paths going through node $i$. Then, the betweenness centrality of node $i$ is
\begin{equation}
    b_i = \sum_{s\neq t} \dfrac{\sigma_i(s, t)}{\sigma(s,t)},
\end{equation}
where we adopt the convention that $\sigma_i(s, t)/\sigma(s,t) = 0$ if both $\sigma_i(s, t)$ and $\sigma(s, t)$ are zero.
Betweenness can be thought of as the amount of load a node must support when there is some kind of flux on the network. Nodes with higher betweenness articulate different groups of nodes and their importance are more related to the communicability of the network.
In recent works, BC has been reported as one of the most powerful strategies for dismantling networks. For example, in~\cite{Iyer2013, Wandelt2018} several models of synthetic networks, as well as real-world networks are studied under different dismantling strategies. In most cases, BC attacks overtake the rest of the strategies considered. In a recent work~\cite{Almeira2020}, it was also reported that even networks with homogeneous betweenness distribution can be efficiently dismantled using betweenness-based attacks. 

As BC is a global measure, its strength as a measure for dismantling processes comes in hand with an important drawback, which is the computational complexity associated with its calculation. 
The most efficient algorithm so far known was proposed by Brandes, \emph{et al.} in~\cite{Brandes2001} and runs in $\mathcal{O}(NM)$, where $N$ and $M$ are the number of nodes and links in the network, respectively. 

Different variations of BC have been proposed, most of which are covered in~\cite{Brandes2008}. In particular, in~\cite{Ercsey-Ravasz2010,Ercsey-Ravasz2012}, Ercsey-Ravasz, et al. study the so-called $\ell$-betweenness centrality, originally proposed by~\cite{Borgatti2006}. The difference with the original definition is that  $\ell$-betweenness ignores paths that are longer than $\ell$. The authors show that the $\ell$-betweenness distributions for different values of $\ell$ can be scaled into a universal curve. In addition, they argue that a moderate value of $\ell$ is sufficient for identifying the influencer nodes, which could give an important improvement in terms of algorithmic complexity, allowing the identification of central nodes in larger systems.
In this work, we argue against this idea by showing that, for large DT networks, the relation between $\ell$ and the network size dramatically affects the nature of the network dismantling process.

%%%%%%%%%%%%%%%%%%%%%%%%%%%%%%%%%%%%%%%%%%%%%%%%%%%%%%%%%%%%%%%%%%%%%%%%%
%%%%%%%%%%%%%%%%%%%%%%%%%%%%%%%%%%%%%%%%%%%%%%%%%%%%%%%%%%%%%%%%%%%%%%%%%
\subsection{Percolation}\label{subsec:Percolation}

 Percolation is a theoretical framework used to describe transitions in which a system changes from a disconnected to a globally connected state and vice versa. For instance, in the case of random site percolation, each node of an existing network is occupied with probability $p$, hence remaining unoccupied with probability $f = 1 - p$, and links are activated only if they connect two occupied nodes.
When $p$ is small, the occupied nodes are apportioned in small-sized components, while if $p$ is larger than a critical threshold $p_c$,  one of the components becomes extensive and the system percolates.           
This extensive component is known as the giant connected component (GCC). Random site percolation generates a continuous transition in the size of the GCC, with critical exponents that are related to the topology of the network. Instead, the fragmentation of networks under targeted attacks can induce abrupt transitions similar to the observed in explosive percolation processes~\cite{DSouza2019}. The nature of these transitions, i.e. whether they are continuous (second-order) or discontinuous (first-order) is not well understood and is a current topic of research. In particular, in the case of betweenness attacks, few works address this point. Previous works in planar graphs~\cite{Norrenbrock2016} and random Erd\"os Renyi networks~\cite{Almeira2020} seem to indicate that recalculated attacks based on betweenness could induce discontinuous transitions. In particular, Norrenbrock, et al.~\cite{Norrenbrock2016} study the percolation transition for both recalculated degree-based (RD) and betweenness-based (RB) attacks on four different models of spatial networks, including DT. They conclude that the RD attack belongs to the standard two-dimensional percolation transition universality class.  For RB, the percolation threshold is located at $f_c =1-p_c = 0$, and even when the nature of the transition was not explored by these authors it seems to be abrupt as in the case of Erd\"os Renyi networks~\cite{Almeira2020}.

%%%%%%%%%%%%%%%%%%%%%%%%%%%%%%%%%%%%%%%%%%%%%%%%%%%%%%%%%%%%%%%%%%%%%%%%%
%%%%%%%%%%%%%%%%%%%%%%%%%%%%%%%%%%%%%%%%%%%%%%%%%%%%%%%%%%%%%%%%%%%%%%%%%
\subsection{Finite-size scaling analysis}

The characterization of the percolation transitions associated with each attack procedure was performed using finite-size scaling analysis (FFSA)~\cite{Cho2010,Fortunato2011}.
According to this theory, the divergence of the correlation length at the percolation threshold implies that every variable of the system becomes scale-independent at that point. A finite-size system of linear size $L:=N^{1/d}$, where $d$ is the number of spatial dimensions, produces a scaling of the form
\begin{equation}
\label{eq:scaling}
    X \sim L^{-\omega/\nu} F[(f-f_c) L^{1/\nu}],
\end{equation}
where $f$ is the fraction of nodes removed, and $\omega$ is an exponent related to the variable $X$. For $f=f_c$, the variable behaves as $X \sim L^{-\omega/\nu}$. 
The relation in Eq.~\ref{eq:scaling} holds asymptotically, i.e. in the limit $L \rightarrow \infty$ and $f \rightarrow f_c$, and it can be used to obtain the ratio $\omega/\nu$ by computing $X(f_c, L)$ for different system sizes. In addition, the plot of $L^{\omega/\nu}X$ as a function of $(f_c-f) L^{1/\nu}$ yields the universal function $F$, which is independent of $L$, so curves corresponding to different sizes collapse. 

An important use case of Eq.~\ref{eq:scaling} is the scaling for the relative size of the components~\cite{Fan2012}
\begin{equation}
\label{eq:comp-scaling}
S_i(f,L) \sim L^{-\beta /\nu} \tilde{S}_i[(f-f_c) L^{1/\nu}].
\end{equation}
Here, the subscript $i=1,2, ...$ indicates the rank of each component, sorted by size in decreasing order. In particular, we will be interested in the order parameter $S_1$ and in the size of the second cluster $S_2L^d$. 
Another useful observable is the second moment of the component size distribution, which is defined as $M_2=\Sigma_s^{'} s^2 n_s$, where $n_s$ is the number of clusters of size $s$ per node and the primed sum excludes the GCC. The scaling relation associated with this metric is
\begin{equation}
\label{eq:suscep-scaling}
M_2(f,L)  \sim L^{\gamma / \nu} \tilde{S} [(f-f_c) L^{1/\nu}].
\end{equation}

For a finite-size system, the percolation threshold does not necessarily coincide with the corresponding value for $N\rightarrow \infty$. In general, the difference between these values presents a scaling in the form
\begin{equation} \label{eq:peak_pos_shift}
f_c(L) - f_c = b L^{-\lambda}.
\end{equation}
In standard percolation, as well as in many other models, $\lambda = 1/\nu$, but this is not always valid~\cite{Li2012,Ziff2010,Grassberger2011ExplosiveBehavior}. For example, in some explosive percolation models, small-sized systems present differences between the two exponents due to large crossover sizes. Also, using \eqref{eq:peak_pos_shift} as a method for estimating both the percolation threshold and exponent $\lambda$ is not recommended if only small sizes and low sampling is available, as it might introduce systematic errors~\cite{Bastas2014}.

In a recently published article, Fan, et al. proposed a new method to analyze generalized percolation processes, based on the scaling of the largest jump in the order parameter during the process~\cite{Fan2020}. Although the authors deal with bond percolation, the same analysis can be performed for site percolation. Based on this work, we define the \emph{gap}
\begin{equation} \label{eq:Delta}
\Delta^{(i)}(L) = \dfrac{1}{L^d} \max_{t} \left[ N^{(i)}_1(t+1) - N^{(i)}_1(t) \right],
\end{equation} 
where $N^{(i)}_1(t)$ is the size of the largest cluster after removing $t$ nodes from network $i$. We also define $t^{(i)}_{\Delta}$ as the number of nodes removed such that the maximum in \eqref{eq:Delta} is attained and the \emph{gap position} $f^{(i)}_{\Delta}(L) = t^{(i)}_{\Delta}/L^d$. This value can be used to estimate the percolation threshold for a finite-size system. Other estimations for this value, such as the peak position of $M_2$ or $S_2L^d$, in general differ from this value, but the differences are expected to decrease as the size of the system increases.

According to~\cite{Fan2020}, the averaged values of the previous quantities present a scaling of the form \footnote{Here, $\Delta(L) := \langle \Delta^{(i)}(L)\rangle$, where the average is taken over realizations $i$ of the attack. $f_{\Delta}(L)$ and $N_{1,\Delta}(L)$ are defined in a similar way.}
\begin{subequations} \label{eq:fan_avg}
    \begin{align}
        \Delta(L) &\sim L^{-\beta/\nu},
        \label{eq:fan_Delta}
        \\
        f_{\Delta}(L) - f_{\Delta}(\infty) &\sim L^{-1/\nu_1}, \label{eq:fan_rc} 
        \\
        N_{1,\Delta}(L) &\sim L^{-d_f}. 
        \label{eq:fan_N1}
    \end{align}
\end{subequations}
Also, the corresponding fluctuations, defined as the standard deviations, scale as
\begin{subequations}\label{eq:fan_chi}
    \begin{align}
        \chi_{\Delta}(L) &\sim L^{-\beta/\nu},
        \label{eq:fan_chiDelta} \\
        \chi_{f_{\Delta}}(L) &\sim L^{-1/\nu},
        \label{eq:fan_chirc} \\
        \chi_{N_{1,\Delta}}(L) &\sim L^{-d_f}. 
        \label{eq:fan_chiN1}
    \end{align}
\end{subequations}
Note that, in a similar way as it happens for Equation \eqref{eq:peak_pos_shift}, the exponent $\nu_1$ does not coincide in general with the correlation length critical exponent $\nu$.

\subsection{Computational and statistical methods}

All simulations and computations were performed using Python and C++, depending on the case. For the construction of DT networks, $N=L^2$ points were drawn uniformly on an $L\times L$ square (uniform density approach) and then the triangulation was computed using the Python package SciPy~\cite{2020SciPy-NMeth}, considering open boundary conditions. Betweenness computation was performed using the python packages igraph~\cite{igraph2} and NetworKit~\cite{Staudt2014NetworKit:Analysis}. We found that the igraph implementation runs faster than NetworKit, but the latter allows parallelization. Percolation analysis was done using our implementation of the Newman-Ziff algorithm~\cite{Newman2001a}, where we introduced a variation that allows us to compute, in linear time, the average finite-cluster size and the size of the second-largest component. Our code is available at the GitHub repository \url{https://github.com/nahuelalmeira/dismantlingScaling}. Averages were computed over $10^2-10^4$ independent networks for each size and attack strategy. Power-law fittings were performed using least-squares linear regression over the logarithm of the corresponding variables. To minimize finite-size effects, fits only include the largest five sizes available. Uncertainties are reported for a confidence interval of $95\%$.

%%%%%%%%%%%%%%%%%%%%%%%%%%%%%%%%%%%%%%%%%%%%%%%%%%%%%%%%%%%%%%%%%%%%%%%%%
%%%%%%%%%%%%%%%%%%%%%%%%%%%%%%%%%%%%%%%%%%%%%%%%%%%%%%%%%%%%%%%%%%%%%%%%%
%%%%%%%%%%%%%%%%%%%%%%%%%%%%%%%%%%%%%%%%%%%%%%%%%%%%%%%%%%%%%%%%%%%%%%%%%
\section{Results}\label{sec:Results}

%%%%%%%%%%%%%%%%%%%%%%%%%%%%%%%%%%%%%%%%%%%%%%%%%%%%%%%%%%%%%%%%%%%%%%%%%
%%%%%%%%%%%%%%%%%%%%%%%%%%%%%%%%%%%%%%%%%%%%%%%%%%%%%%%%%%%%%%%%%%%%%%%%%
\subsection{Full-range betweenness}\label{subsec:full-range}

\begin{figure}[t]
\centering
\includegraphics[scale=0.28]{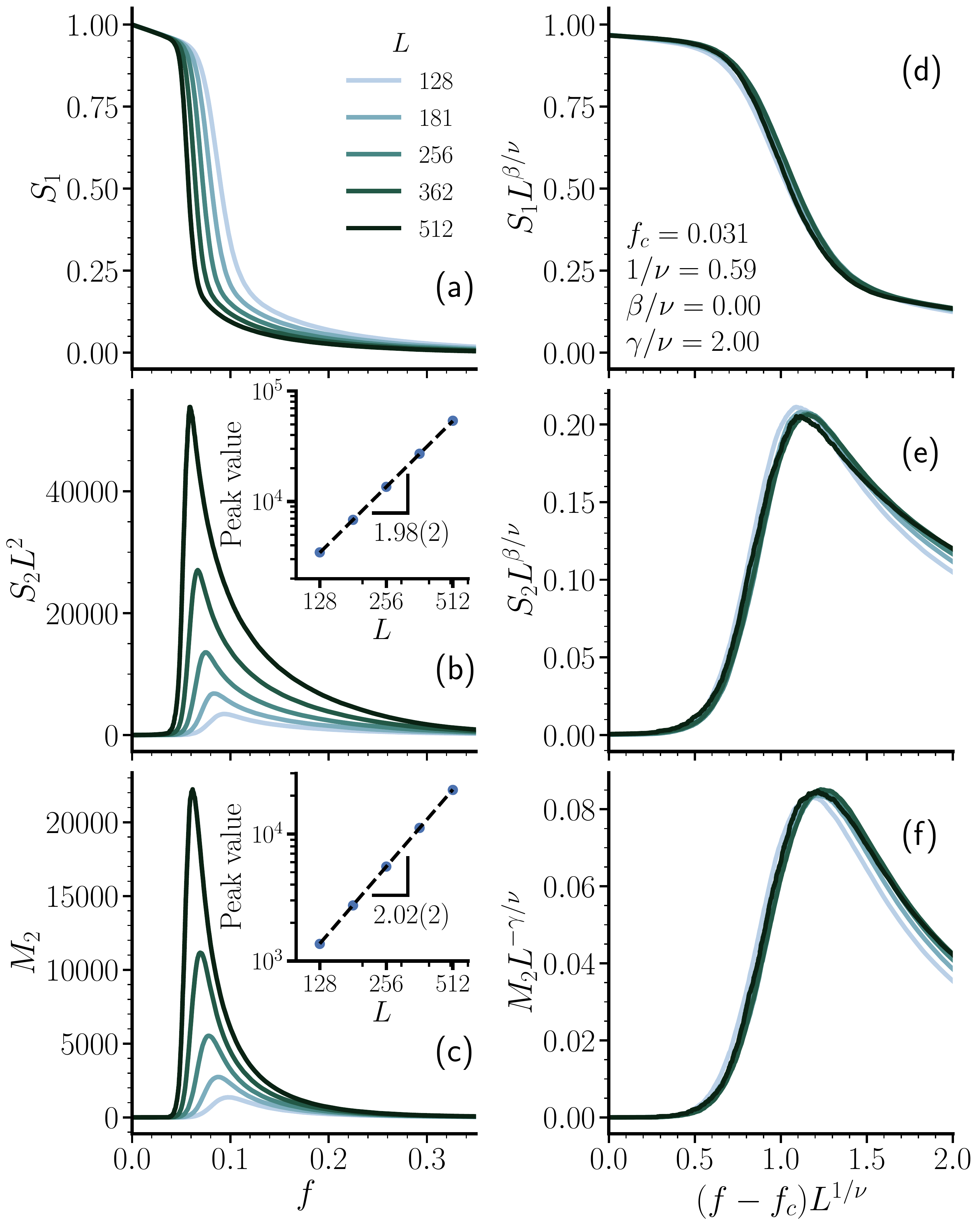}
\caption{\label{fig:collapse_B} \textbf{(a-c)} Relative size of the giant component $S_1$, size of the second-largest cluster $S_2L^2$, and second moment of the finite-cluster size distribution $M_2$ as a function of the fraction of nodes removed for the full-range betweenness attack. 
Insets in Panels (b-c) show the scaling of the peaks for  $M_2$ and $S_2 L^2$ with the system size. The exponent ratios obtained from the fit of the peaks are $1-\beta/\nu = 1.98(2)$, and $\gamma/\nu=2.02(2)$, which are consistent with a first-order transition ($\beta = 0$ and $\gamma = 2\nu$).
\textbf{(d-f)} Collapse of the curves from left panels based on Eqs. \eqref{eq:comp-scaling} and \eqref{eq:suscep-scaling}.}
\end{figure}

Figures \ref{fig:collapse_B}b and \ref{fig:collapse_B}c show the second moment of the finite-size distribution $S_2L^2$ and the size of the second-largest cluster $M_2$. Both metrics peak close to the percolation threshold. Following the scaling ansatz \eqref{eq:comp-scaling} and \eqref{eq:suscep-scaling}, we computed the exponent ratios $\beta / \nu$ and $\gamma / \nu$ by fitting the peak sizes of $S_2L^2$ and $M_2$ for different system sizes (see figure insets). From the peak height of the second-largest cluster, we obtained $\beta/\nu = 0.02(2)$. Similarly, a ratio $\gamma/\nu = 2.02(2)$ was obtained from the peak of the second moment. 
The latter exponent ratio can be also obtained from the scaling of the fluctuations of the order parameter $\chi = \sqrt{\langle S_1^2\rangle - \langle S_1\rangle^2}$. 
We present the analysis in the Supplementary Material (\cite{SM}, Section S1), by which an estimation $\gamma/\nu=2.04(2)$ was obtained. The critical exponents found for this attack strategy are consistent with a first-order phase transition ($\beta = 0$ and $\gamma = d\nu$~\cite{Binder1981,Binder1984, Araujo2010}). 

The right panels of Figure \ref{fig:collapse_B} show the collapse of each percolation metric based on the scaling ansatz \eqref{eq:comp-scaling} and \eqref{eq:suscep-scaling}. We have chosen the parameters that give the best collapse, taking into account that they could slightly differ from the parameters estimated by other means, especially when small sizes are considered. 

\begin{figure}[ht]
\centering
\includegraphics[scale=0.28]{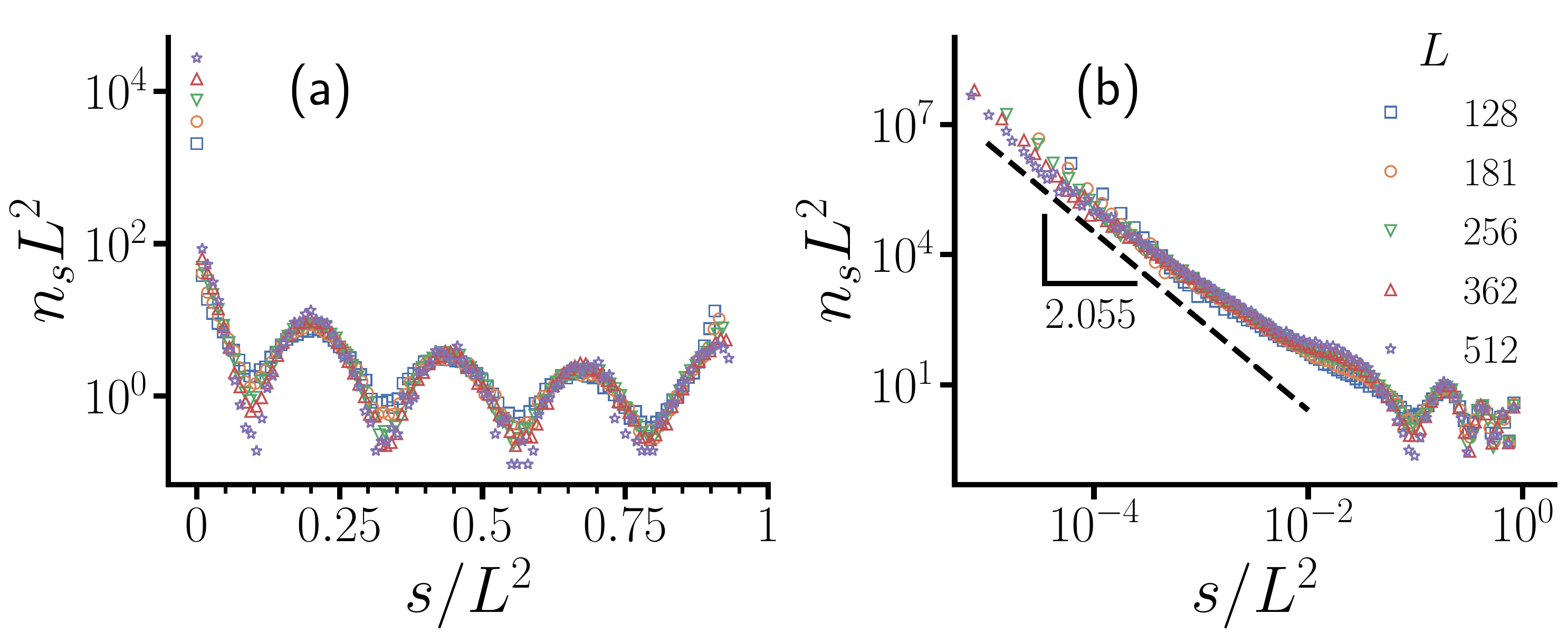}
\caption{\label{fig:comp_sizes_B} Scaled component size distribution for the full-range betweenness attack at $f=f_{\Delta}(L)$, including the largest cluster.   Each histogram is built combining $10^4$ networks. 
\textbf{(a)} Linear binning, showing a series of peaks and valleys that follow a characteristic frequency. 
\textbf{(b)} Logarithmic binning, exposing an heterogeneous distribution of small sizes. The dashed line corresponds to the expected distribution for random percolation on an infinite two-dimensional lattice.
}
\end{figure}

We also analyzed the component size distribution close to the percolation threshold for the full-range betweenness attack. Instead of considering only the finite-size components, we include the giant component, as it can give insights regarding the nature of the transition. In Figure \ref{fig:comp_sizes_B} we show histograms for this distribution (aggregated over $10^4$ independent simulations) for different system sizes at the gap position $f_{\Delta}(L)$. 
In Panel \ref{fig:comp_sizes_B}a, where a linear binning is employed, we see a multimodal distribution with peaks that coincide when data are scaled by the number of nodes $L^2$. We note that the separation between peaks is roughly equal, and approximately 0.25. As we will discuss later, the location of the peaks can be understood by examining in detail the attack evolution in single realizations. The logarithmic binning in Panel \ref{fig:comp_sizes_B}b shows that the distribution stretches for lower sizes. %The slope at this part of the distribution is smaller than 

%%%%%%%%%%%%%%%%%%%%%%%%%%%%%%%%%%%%%%%%%%%%%%%%%%%%%%%%%%%%%%%%%%%%%%%%%
%%%%%%%%%%%%%%%%%%%%%%%%%%%%%%%%%%%%%%%%%%%%%%%%%%%%%%%%%%%%%%%%%%%%%%%%%
\subsubsection*{Largest gap statistics}

\begin{figure}[t]
\centering
\includegraphics[scale=0.28]{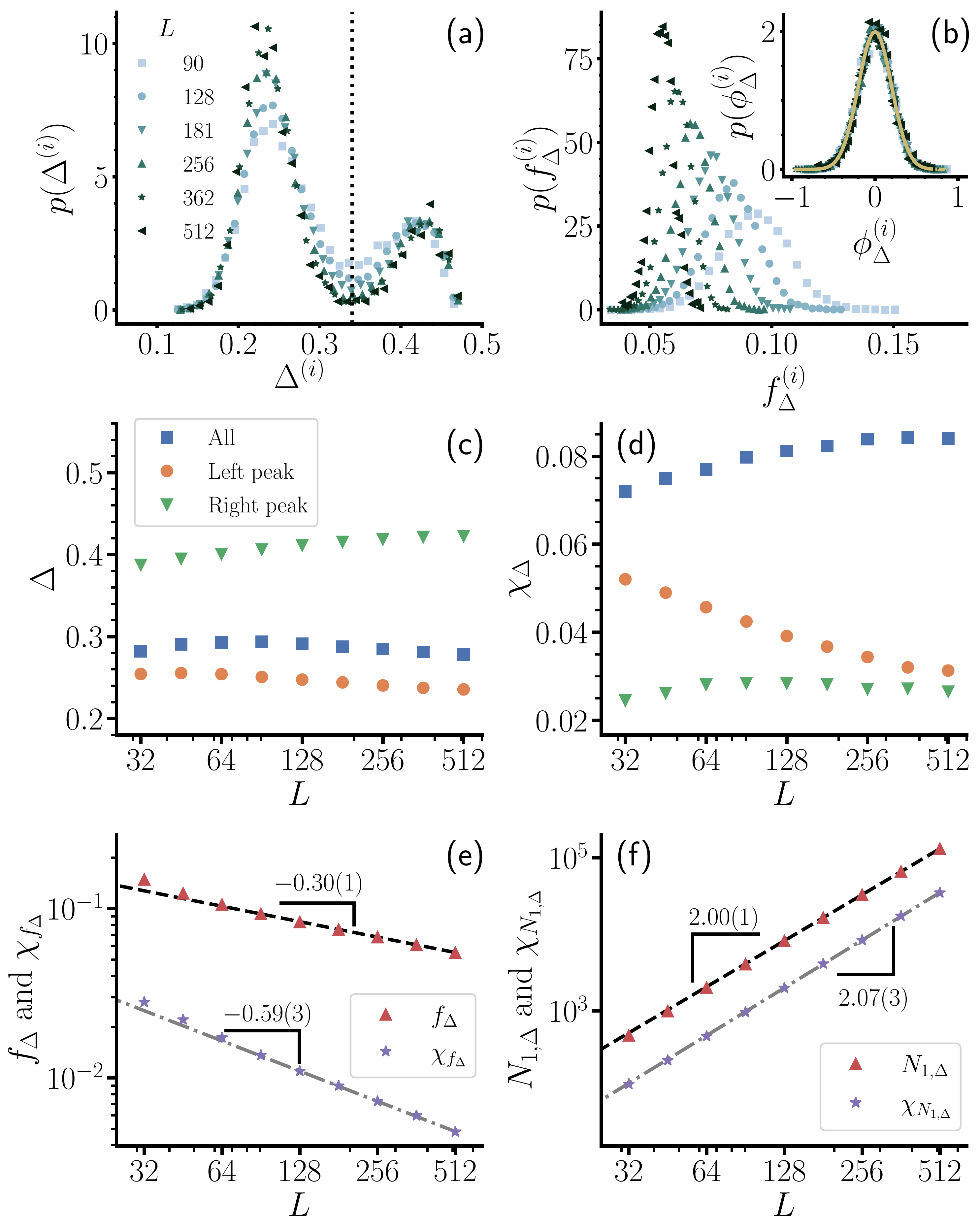}
\caption{
\label{fig:gap_exponent_B} 
Largest one-step gap statistics for the full-range betweenness attack as a function of the network size.
\textbf{(a)} Probability distribution for the gap size. As the system size increases, the distribution becomes bimodal and the values between the two modes become less probable. 
\textbf{(b)} Probability distribution for the gap position. The curves shift towards the left and become more peaked as $L$ increases. 
Inset: Collapse of the histograms after the change of variables $\phi^{(i)} = (f_{\Delta}^{(i)} - f_{\Delta})L^{1/\nu}$. The corresponding master curve (yellow curve) is a Gaussian distribution.
\textbf{(c-d)} Average and standard deviation of the gap size. Blue squares include all simulations, while orange circles and green triangles discriminate the values corresponding to the left and right peaks of the probability distribution, respectively.
\textbf{(e)} Scaling for the average gap position $f_{\Delta}$ and its fluctuations $\chi_{f_\Delta}$. From the scaling of $f_{\Delta}$, it can be seen that the largest gap occurs, in the thermodynamic limit, at the beginning of the attack. From the scaling of the fluctuations, we can estimate the correlation length exponent $1/\nu = 0.59(3)$. \textbf{(f)} Average size and fluctuations for the giant component at the gap position. The scaling of these two metrics indicates a fractal dimension of the percolating cluster of $d_f = 2$, consistent with a first-order transition.}
\end{figure}

We complement the previous analysis by studying the largest gap statistics similarly as in~\cite{Fan2020}. Our main results are summarized in Figure  \ref{fig:gap_exponent_B}. In Panel  \ref{fig:gap_exponent_B}a we plot the probability distribution of gap sizes, computed over $10^4$ independent simulations. The distributions are bimodal, with peaks that become sharper as the system size increases. The height of the right peak does not seem to depend on $L$, but the left peak increases as the system size increases. 
The positions of the peaks, which correspond to the typical largest one-step damage produced by the attack, remain constant at about $0.25$ and $0.45$. We will give an interpretation of these values in the following section, based on a geometrical characterization of the attack.

Panel \ref{fig:gap_exponent_B}b shows the corresponding probability distribution for the position of the largest gap. Contrary to the gap size, where two typical values are observed, the gap position exhibits a centered, unimodal distribution. As the system size increases, the curves shift towards the left and become sharper. Based on the arguments presented in~\cite{Fan2020}, we define the variable $\phi^{(i)} = (f_{\Delta}^{(i)} - f_{\Delta})L^{1/\nu}$, where $\nu$ is the correlation length exponent and plot its probability distribution for each system size (see panel inset). The curves for all sizes collapse well into a master curve (yellow curve) which, through the central limit theorem, corresponds to a Gaussian distribution.

%Let us bring the discussion again to the statistics of the gap size.
As it was introduced in Section \ref{sec:Methods}, for continuous percolation the average gap size $\Delta(L)$ vanishes in the thermodynamic limit, following a power-law with associated exponent $-\beta / \nu$. First-order transitions, in turn, are typically characterized by $\beta = 0$ \footnote{An exception to this rule are the so-called hybrid percolation transitions, which describe some explosive percolation models \cite{DSouza2015}.} and thus their average gap size remains finite even for $L\rightarrow \infty$. 
Our data shows consistency with the latter case, as we show in Panel \ref{fig:gap_exponent_B}c. When all simulations are considered (blue squares), the average gap size remains approximately constant, with a slight drop for larger systems. However, we note that an average computed over a bimodal distribution could be misleading, so we also computed averages restricted to each of the two peaks of $p(\Delta^{(i)})$ (separated by the vertical dashed line in Panel \ref{fig:gap_exponent_B}a). When doing this, we see that each average approximates to the modes as the system size increases.

Alongside the averages, the fluctuations of the gap size are expected to present the same scaling properties (see \eqref{eq:fan_chi}). When computed over all simulations, we see an increment of the fluctuations for smaller systems, but the values seem to stabilize for larger networks. This increment is associated with the concentration of values close to the peaks in $p(\Delta^{(i)})$, which is more notorious for smaller values of $L$. If we discriminate each peak, as we did for the averages, the fluctuations converge to approximately the same value. 

We now discuss the average and fluctuations of the gap position. As seen in Panel \ref{fig:gap_exponent_B}e, equation \eqref{eq:fan_rc} is satisfied with a scaling exponent $1/\nu_1 = 0.30(1)$ and thermodynamic gap position $f_{\Delta}(\infty) = 0$. That is, the largest one-step damage of the attack occurs when a sub-extensive fraction of nodes are removed. In terms of network robustness, this implies that the betweenness-based strategy is extremely efficient in dismantling Delaunay triangulations. As for the fluctuations of the gap position, the power-law relation \eqref{eq:fan_chirc} is satisfied with inverse correlation-length exponent $1/\nu=0.59(3)$. This is the value we employed for collapsing the curves in Figure \ref{fig:collapse_B}, and in Panel \ref{fig:gap_exponent_B}b.

To conclude this section, we analyze the average size of the largest cluster at the gap position $N_{1,\Delta}$ and its corresponding fluctuations $\chi_{N_{1,\Delta}}$. We see from Panel \ref{fig:gap_exponent_B}f that both quantities scale with exponents that are consistent with $d_f = 2$. In other words, the scaling predicts that the percolation cluster is not a fractal, but a regular two-dimensional object. If the hyperscaling relation $d-d_f = \beta/\nu$ holds, then $\beta = 0$ and thus, the transition is discontinuous.

%%%%%%%%%%%%%%%%%%%%%%%%%%%%%%%%%%%%%%%%%%%%%%%%%%%%%%%%%%%%%%%%%%%%%%%%%
%%%%%%%%%%%%%%%%%%%%%%%%%%%%%%%%%%%%%%%%%%%%%%%%%%%%%%%%%%%%%%%%%%%%%%%%%
\subsubsection*{Geometrical characterization of the attack}

\begin{figure*}[t]
\centering
\includegraphics[scale=0.24]{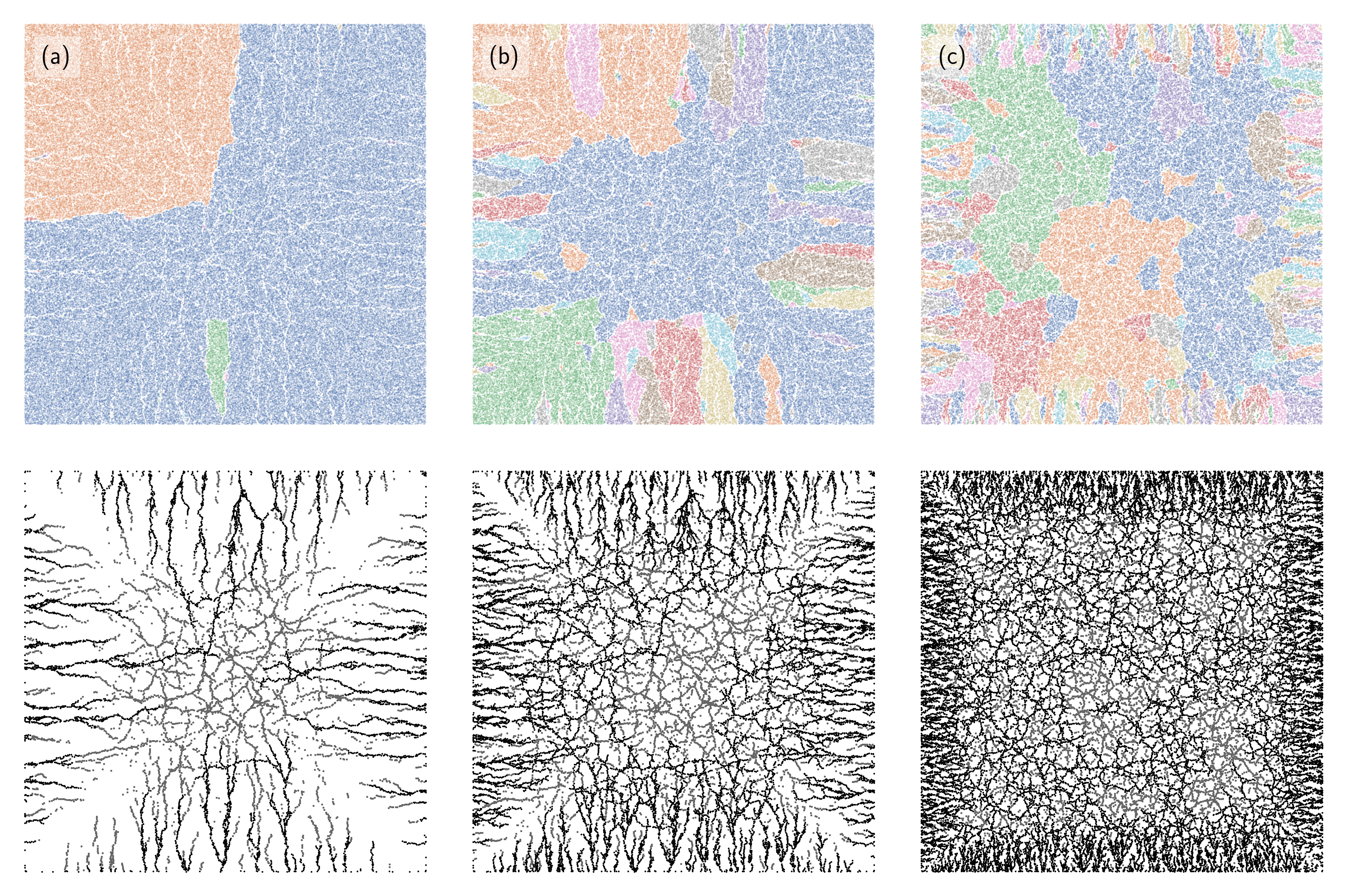}
\caption{\label{fig:breaking}
Single realization of different attacks on a DT network with size $L=512$ at $f=f_{\Delta}$. Colored nodes in the upper panels correspond to different connected components. Giant and second-largest clusters are colored blue and orange, respectively. Nodes that have been removed are plotted as black stars and gray squares in the corresponding lower panels. The former belong to the GCC of the subgraph induced by the removed nodes. \textbf{(a)} Full-range betweenness attack. After removing the braking nodes, the network splits into two extensive clusters plus, eventually, small-size components. \textbf{(b-c)} Range-limited betweenness attack with $\ell=128$ and $\ell = 64$. As the cutoff diminishes, the behavior of the transition changes and a broad distribution of component sizes emerges, progressively approaching a continuous phase transition.
}
\end{figure*}

To gain more intuition about the nature of the transition, it is useful to look at a single realization of the attack procedure. 
Also, to reinforce ideas, we perform a comparative analysis by using range-limited betweenness attacks as we will explain in the next sections.
As an example, we show in Figure \ref{fig:breaking} the state of a network for a realization $i$ at $f=f^{(i)}_{\Delta}$. In other words, we present the network one step after the occurrence of the largest gap. The size of the network is $L=512$. Each panel represents a different attack strategy. In particular, Panel \ref{fig:breaking}a corresponds to the full-range betweenness attack. To keep the figure clean, edges are omitted and only nodes are drawn. Colored points correspond to the nodes that remain in the network, each color representing a connected component. On the other hand, gray and black nodes are those that have been removed ---i.e., the nodes with the highest betweenness. The difference between gray and black nodes is that the black ones form the largest connected component of the subgraph induced by the removed nodes. 

The first thing to notice is that the high betweenness nodes are located either in the central part of the network or over striations that go from the center to the periphery. Due to the symmetry of the network, these paths are mainly horizontal or vertical. As a consequence of the open boundary conditions, the nodes located on the border are connected by longer edges than bulk nodes. For this reason, they tend to have large betweenness. These nodes also connect different striations to the largest component of removed nodes.

As it can be inferred from the figure, the largest break in the network occurs when two of these paths, coming from different borders, merge. As a consequence, a fragment of about a quarter of the network (if the paths come from adjacent borders) or half of the network (for paths coming from opposite borders) is separated from the giant component. This fact explains the bimodal distribution observed for the largest gap size $\Delta$, and the position of the peaks for the component size distribution, located roughly at $s/L^2=1/4, 1/2, \mathrm{\;and\;} 3/4$.

We define the set of nodes removed up to $f=f_{\Delta}$ as the \emph{vulnerability backbone} of the network \footnote{A similar definition is given in~\cite{Ercsey-Ravasz2010}.}. Our definition is based on two main facts. First, the removal of these nodes produces a massive breakdown of the network and second, the amount of nodes $t_{\Delta}$ included in this set is a sub-extensive quantity, as it scales as $t_{\Delta}(L) = L^2 f_{\Delta}(L)\sim L^{2-1/\nu_1}$, with $1/\nu_1=0.30(1)$.

%%%%%%%%%%%%%%%%%%%%%%%%%%%%%%%%%%%%%%%%%%%%%%%%%%%%%%%%%%%%%%%%%%%%%%%%%
%%%%%%%%%%%%%%%%%%%%%%%%%%%%%%%%%%%%%%%%%%%%%%%%%%%%%%%%%%%%%%%%%%%%%%%%%
%%%%%%%%%%%%%%%%%%%%%%%%%%%%%%%%%%%%%%%%%%%%%%%%%%%%%%%%%%%%%%%%%%%%%%%%%
\subsection{Range-limited betweenness}\label{subsec:range-limited}

As we have discussed so far, betweenness centrality turns out to be a useful metric for assessing the vulnerability of random lattices. This effectiveness is probably associated with the fact that betweenness is a global measure, retrieving information from the whole network for each node. One way we can test this assumption is by introducing a parameter to the definition of betweenness, to tune the range of the interactions. We do this by employing $\ell$-betweenness and by performing a systematic study of the percolation transition in terms of the cutoff value $\ell$.

\begin{figure}[ht]
\centering
\includegraphics[scale=0.28]{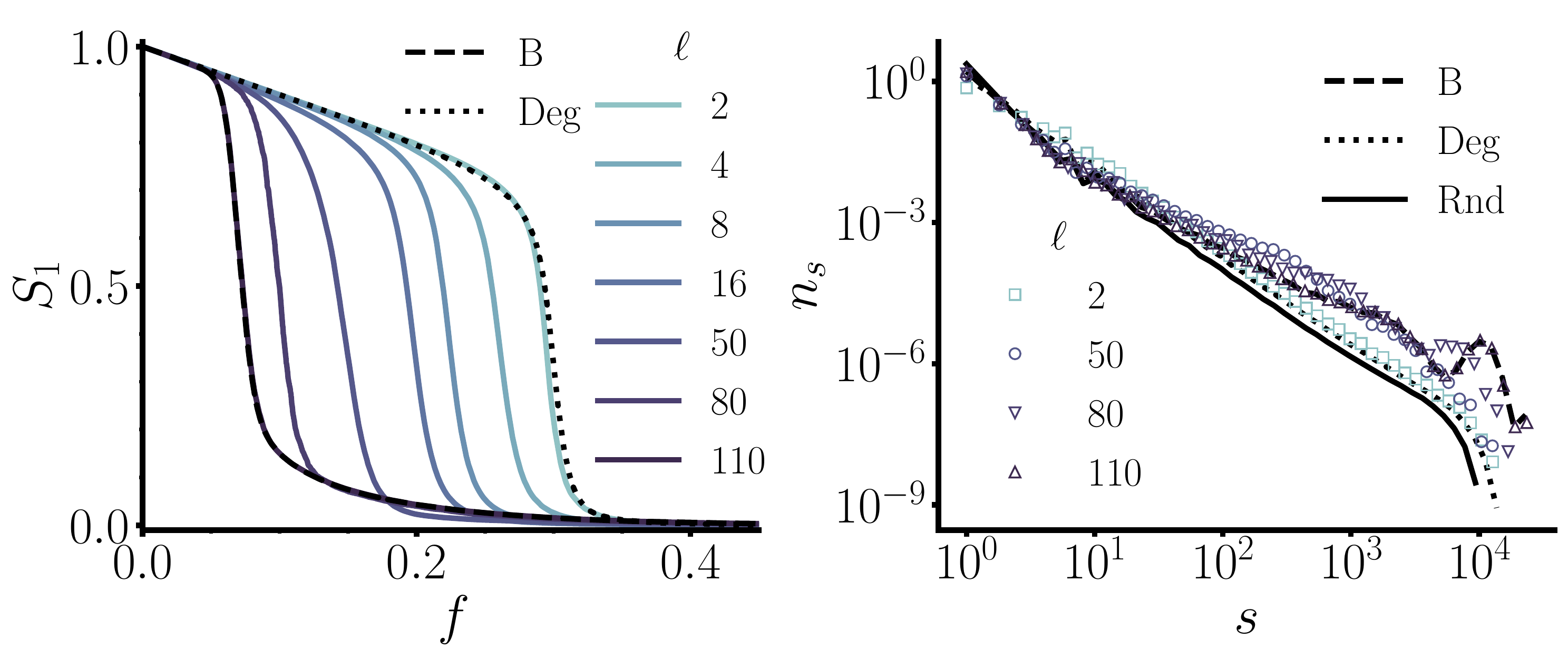}
\caption{\label{fig:Bl_attacks} Characterization of range-limited betweenness attacks for different cutoff values on DT networks of linear size $L=256$. 
\textbf{(a)} Evolution of the order parameter $S_1$. For $\ell=2$, the attack behaves in a similar fashion as the degree-based attack (dotted line). As $\ell$ increases, the percolation threshold shifts towards the left until, for $\ell\approx 110$, the attack becomes indistinguishable from the full-range attack (dashed line).
\textbf{(b)} Component size distribution $n_s$ at the percolation threshold for range-limited attacks with different cutoffs. The distribution excludes the giant component and the percolation threshold is estimated as the position where $M_2$ peaks. For small cutoffs, the behavior is similar to random percolation (solid black line). As $\ell$ increases, the curves flatten and a bump for large values of $s$ emerges.}
\end{figure}

In Figure \ref{fig:Bl_attacks}a we show the evolution of the order parameter $S_1$ as a function of the fraction of removed nodes employing attacks for several values of $\ell$ on networks with linear size $L=256$.
Curves are averaged over $10^3-10^4$ realizations, with errors lower than line width. As expected, a more efficient dismantling process is observed as the cutoff length increases. In addition, the associated percolation transition becomes sharper for larger values of $\ell$. 
Eventually, we see that for a large enough cutoff the range-limited attack cannot be distinguished from the full-range version. This suggests the existence of a crossover cutoff $\ell^*(L)$ such that $\mathrm{B}\ell$ attacks perform as well as $\mathrm{B}$ for all $\ell>\ell^*$.

An alternative comparison between the different attacks can be made by studying the finite-component size distribution $n_s$ at the critical point (see Figure  \ref{fig:Bl_attacks}b). Before discussing the results, we point out that the quantities plotted in this figure differ from the ones presented in Figure \ref{fig:comp_sizes_B}b in two ways. First, the distribution shown in  \ref{fig:Bl_attacks}b does not include the giant component, and second, instead of $f_{\Delta}(L)$, we choose the position of the peak of the second moment $M_2$ as the estimator for the finite-size percolation threshold $f_c(L)$.
For attacks with a short cutoff value, it is expected that the finite-component size distribution does not differ significantly from standard percolation. As the figure shows, this is indeed the case. For larger values of $\ell$, in turn, clear differences can be observed.
Specifically, the distribution flattens and a bump for larger sizes arises. As we have excluded the giant component in this calculation, we do not observe the three peaks shown in Figure \ref{fig:comp_sizes_B} for the full-range attack. Nevertheless, the bump at the end of the distribution for large cutoff values points towards the presence of a sharp transition, acting as a power-keg in a similar way as in other explosive percolation models~\cite{Friedman2009}.

To complement the former qualitative description of the attacks 
%based on studying networks of a given size
, we performed a finite-size scaling analysis to characterize the thermodynamic properties of the associated percolation transitions. To keep the section simple, we discuss here only the main results and refer the reader to the Supplementary Material~\cite{SM} for a detailed description of the methods employed. 
On one side, we computed the percolation threshold $f_c^{\mathrm{B}\ell}$ for different values of $\ell$, up to $\ell=16$. Consistently with Figure \ref{fig:Bl_attacks}, the threshold diminishes as the cutoff increases. 
An interesting point here would be to know the value to which the succession of range-limited percolation thresholds converges as $\ell$ grows. That is, $f_c^{\mathrm{B}\infty} = \lim_{\ell\rightarrow \infty} f_c^{\mathrm{B}\ell}$, where the limit $\ell\rightarrow \infty$ is taken after the limit $L\rightarrow \infty$. One possibility is that $f_c^{\mathrm{B}\infty}=f_c^{\mathrm{B}}$, which we have determined as being zero. %Nonetheless, there is not a fundamental reason for which these two values should coincide and, rather, our results suggest that indeed $f_c^{\mathrm{B}\ell}$ could converge to a value strictly larger than zero. To confirm our finding, larger values of $\ell$ should be considered, which is difficult to attain given the computational cost involved in the calculations ---not only the cost of $\ell$-betweenness increases significantly, but also larger networks have to be used to avoid finite-size effects.
Nonetheless, there is not a fundamental reason for which these two values should coincide, and later in the section, we discuss evidence supporting that the limit is indeed strictly larger than zero. From the results discussed so far it is not possible to extrapolate to $\ell\rightarrow \infty$, as the largest $\ell$ for considered was 16, which is a relatively small value. Larger cutoffs are hard to analyze, given the computational cost involved in the calculations ---not only does the cost of $\ell$-betweenness increase significantly but also larger networks have to be used to avoid finite-size effects.

To complete the characterization of the transitions, we computed the critical exponent ratios $\gamma/\nu$ and $\beta/\nu$ employing the same methods as for the full-range betweenness. The values obtained for different cutoffs in the range $2 \leq q \leq 16$ are all consistent with standard percolation on two dimensions. This indicates that, although the transitions occur sooner as $\ell$  increases, they all obey the same underlying dynamics and thus, belong to the same universality class.

\begin{figure}[ht]
\centering
\includegraphics[scale=0.35]{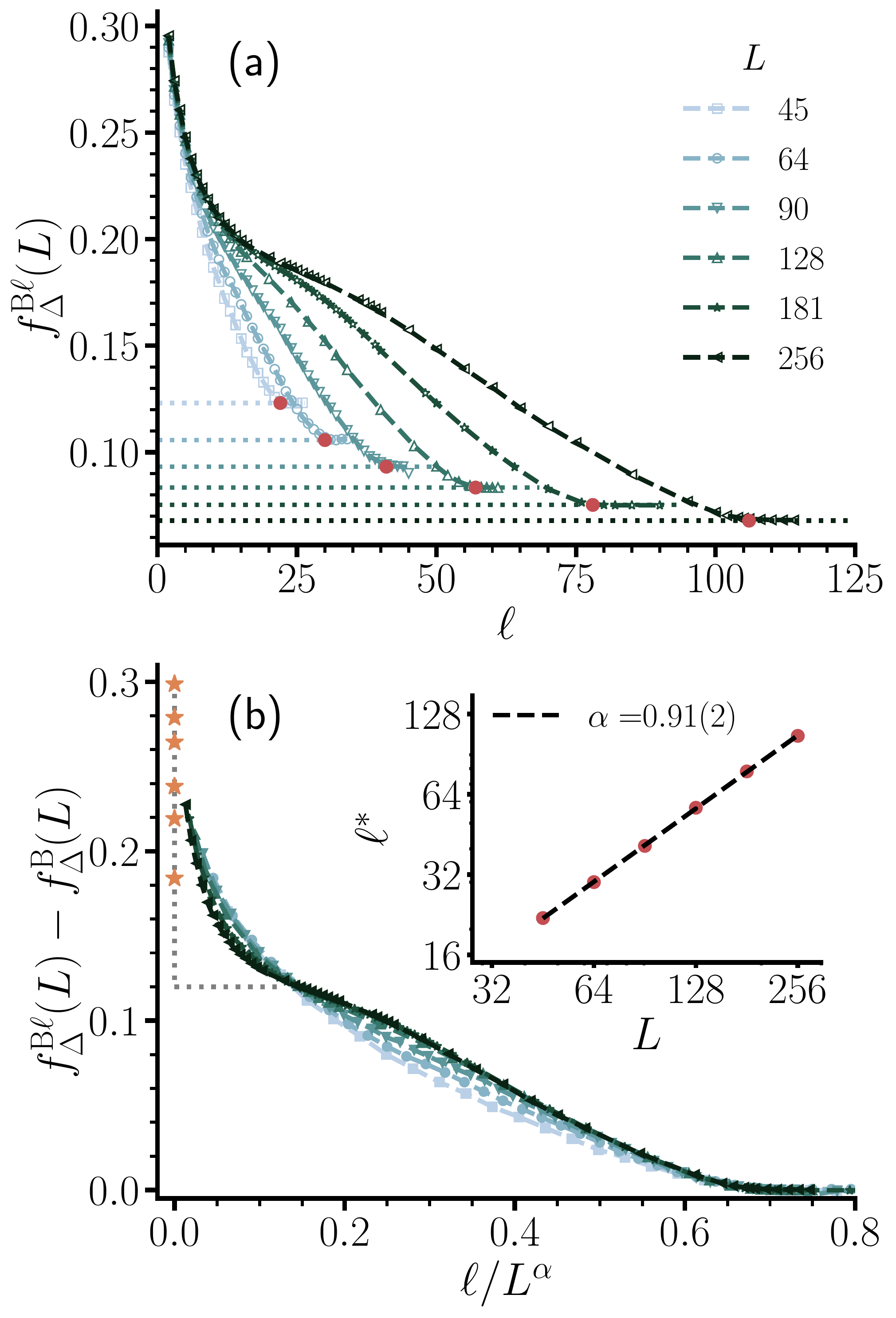}
\caption{\label{fig:phase_diagram} \textbf{(a)} Largest gap position $f_{\Delta}^{\mathrm{B}\ell}(L)$ as a function of the cutoff length $\ell$ for different system sizes. For each value, the position drops until it reaches the value corresponding to the full-range attack (dotted-lines) at $\ell=\ell^*(L)$, marked as red dots. The scaling of $\ell^*$ with $L$ is plotted in the inset of the next panel. \textbf{(b)} The same data from panel (a) presented in scaled coordinates. The vertical axis is shifted by subtracting the largest gap position corresponding to the full-range attack, and the horizontal axis is scaled by $L^{\alpha}$. Orange stars indicate the percolation thresholds $f_c^{\mathrm{B}\ell}$ for $\ell = 2, 3, 4, 6, 8, \mathrm{ and }\;16$.}
\end{figure}

As discussed at the beginning of the section, the behavior of the full-range betweenness attack can be recovered if large enough cutoffs are employed. To give a precise value of $\ell^*(L)$ for each system size, we computed the location of the largest gap $f_{\Delta}$ for different cutoff lengths and system sizes (Figure \ref{fig:phase_diagram}a).
The curves decrease monotonically as $\ell$ increases until they reach the value corresponding to the full-range attack (horizontal dashed lines). We define the crossover cutoff $\ell^*$ as the lowest $\ell$ such that the relative difference between $f_{\Delta}^{\mathrm{B}\ell}$ and $f_{\Delta}^{\mathrm{B}}$ does not exceed a given threshold $c$, and take $c = 0.01$ (the results are not sensitive to the specific value of the threshold, as long as it remains small). The inset shows that the crossover cutoff scales as $\ell^*\sim L^{\alpha}$, with $\alpha = 0.91(2)$.

In addition to the scaling of $\ell^*$, more information can be extracted from the behavior of $f_{\Delta}^{\mathrm{B}\ell}(L)$. As can be seen in Panel  \ref{fig:phase_diagram}a, the largest drop in the gap position occurs for $\ell<10$ independently of the system size. After this approximate value, the shape of the curves becomes more dependent on the system size. While smaller networks continue with a significant drop until they reach  $f_{\Delta}^{\mathrm{B}}$, larger systems exhibit a plateau, followed by an inflection point. Thus, for systems large enough moderate values of $\ell$ do not seem to add too much relevant information regarding node centrality. This might be explained by the fact that the network model here studied does not have mesoscopic structures such as communities which could be exploited by moderate-ranged centrality measures to find weak spots. The information required to dismantle the network is thus encoded %either in the neighborhood of each node, or in the network as a whole. 
at all scales, ranging from each node's neighborhood to the whole system.

Panel \ref{fig:phase_diagram}b shows the curves scaled by $L^{\alpha}$, and shifted vertically by subtracting $f_{\Delta}^{\mathrm{B}\ell}(L)$. The collapse around $\ell/L^{\alpha}\sim 0.7$ indicates that the criterion taken as a definition for $\ell^*$ makes sense (that is, it does not depend significantly on the threshold chosen). In the thermodynamic limit, 
%$f^{\mathrm{B}}_{\Delta}(\infty) = 0$ and $f^{\mathrm{B}\ell}_{\Delta}(\infty) = f_c^{\ell}$ 
this plot corresponds to the phase diagram of the model, which depends on the intensive parameters $f$ and $\ell/L^{\alpha}$. Although it is not possible to know the exact shape of the curve that separates the two phases, we can sketch it by extrapolating the finite-size curves. 
The first thing to notice is that, above a certain system size, all curves cross at a single point $\ell/L^{\alpha} = a\simeq 0.12$. The existence of such a crossing point is an indicator that a non-trivial phase diagram exists. 
Moreover, if we look at the region $\ell/L^{\alpha} > a$, the curves tend to accumulate as the system size increases. We can infer from this fact that the curve at the thermodynamic limit will not differ significantly from the curve corresponding to $L=256$ (the largest size studied). On the contrary, for $\ell/L^{\alpha} < a$ the curves seem to move consistently to the left. We hypothesize that, in the thermodynamic limit, the curve will reach the vertical axis on a finite value $f_c^*$, as sketched by the dotted gray line. For this to happen, the percolation thresholds $f_c^{\mathrm{B}\ell}$ should converge to a non-zero value and in that case, $f_c^* = f^{\mathrm{B}\infty}_c$ ---see previous discussion on the possibility of $f_c^{\mathrm{B}\infty}$ being larger than zero. For comparison, the percolation thresholds $f_c^{\mathrm{B}\ell}$ for $\ell = 2, 3, 4, 6, 8, \mathrm{ and }\;16$ are shown (orange stars).

To complete this section, we resume the discussion on the geometrical characterization of the attacks. Alongside the full-range betweenness attack,
Figure \ref{fig:breaking} shows the same network at $f=f_{\Delta}^{\mathrm{B}\ell}(L)$, for $\ell=128$ (Panel b) and $\ell=64$ (Panel c). Both cutoffs are under the crossover 
value $\ell^*(L)$, and the differences to the full-range attack are evident. If we look at the connected components, we see a picture that is more familiar to standard percolation ---there is a large cluster, followed by multiple finite-size components of heterogeneous sizes. Also, the spatial distribution of the removed nodes changes significantly. We see that the striations do not meet the center, but terminate approximately at a distance $\ell$ from the border. We can use this observation to make a heuristic prediction of $\ell^*$ in the following way. For a massive breakdown of the network to happen, striations coming from different  borders must meet at the center. If the length of these paths is approximately $\ell$, then  $\ell^*(L)\approx L/2$.
%{\color{magenta}[JUAN: quiz\'as el $\alpha<1$ es porque la fase que nuclea en el centro tiene un borde fractal. Ser\'ia interesante medir la dimensi\'on fractal de dicho borde en la Fig.~4c o la 4f.]} {\color{orange} [Nahuel: Eso no daría un $\alpha > 1$?]}
If we recall the values obtained for the crossover cutoff (inset of Figure \ref{fig:phase_diagram}), we can see that up to the sizes studied, the values are close to this estimation.

%%%%%%%%%%%%%%%%%%%%%%%%%%%%%%%%%%%%%%%%%%%%%%%%%%%%%%%%%%%%%%%%%%%%%%%%%
%%%%%%%%%%%%%%%%%%%%%%%%%%%%%%%%%%%%%%%%%%%%%%%%%%%%%%%%%%%%%%%%%%%%%%%%%
%%%%%%%%%%%%%%%%%%%%%%%%%%%%%%%%%%%%%%%%%%%%%%%%%%%%%%%%%%%%%%%%%%%%%%%%%
\section{Discussion}\label{sec:Discussion}

In~\cite{Ercsey-Ravasz2010, Ercsey-Ravasz2012}, Ercsey-Ravasz, et al. perform a systematic analysis on the contribution to betweenness centrality coming from geodesics of different lengths. The authors show that $\ell$-betweenness obeys a characteristic scaling versus $\ell$ which allows to accurately approximate the  full-range betweenness distribution without the expensive computation of long paths. 
As a practical example, they apply their methods to the identification of top-betweenness nodes in different network architectures. In this sense, our systematic analysis of range-limited attacks exposes a limitation of this procedure. Even if the majority of the top-ranked nodes are correctly identified, the thermodynamic aspects of the $\ell$-betweenness attacks are qualitatively different from their full-range counterpart. This suggests that the topological interaction of nodes at a global scale can give important contributions to betweenness, which are missed when local approximations are employed.
We also note that the algorithm presented by the authors does not reduce the algorithmic complexity involved in the calculation of different $\ell$-betweenness centrality --see Supplementary Material~\cite{SM} for performance comparison.
Closely related to the former, we point out another contribution of our work. The thermodynamic characterization of percolation transitions under attacks based on related centrality measures can expose differences between these measures that are not evident from other analyses. Thus, the dismantling approach can be used to build benchmarks for comparing different centrality measures.

In the statistical physics literature, phase transitions on Delaunay triangulations have been extensively studied \cite{Janke1995Two-dimensionalStudy, Lima2000CriticalLattice, DeOliveira2016ContinuousLattices, McCarthy1987InvasionLattice}. Together with bond dilution and local rewiring, the DT represents a model of spatially embedded systems with topological or quenched disorder \cite{Janke2004Harris-LuckLattices, Barghathi2014PhaseDisorder}. One of the consequences of this disorder is a rounding effect on phase transitions. In some systems, disorder destroys phase coexistence and has a rounding effect on first-order transitions \cite{Cardy1999QuenchedTransitions}. For the particular case of full-range betweenness attack, the opposite occurs. If we consider a lattice with periodic boundary conditions, all nodes are equivalent and thus, any attack based on initial centrality is equivalent to random percolation. The disorder introduced by the DT graph sharpens the transition, transforming it from second- to first-order\footnote{To be fair, the comparison should be made with respect to a DT networks with periodic boundary conditions. Preliminary work (not shown) indicates that the transition is not sensitive to the change of boundary conditions.}.

As a possible application of our work to real-world networks, we recall that DT networks have been employed as a model for cities~\cite{Kirkley2018}, where betweenness centrality has a natural interpretation in terms of traffic flows. In this context, high-betweenness node deletion can be interpreted as the saturation of high traffic intersections. Both the discontinuity of the transition and its location at $f=0$ are indicators of the potential fragility of traffic networks. In fact, discontinuous percolation transitions in real-world traffic networks have been recently observed~\cite{GuanwenZengabJianxiGaocLouisShekhtmandShengminGuoeWeifengLveJianjunWufHaoLiugOrrLevydDaqingLiab1ZiyouGaoh1H.EugeneStanleyij12019}. Thus, we believe our work could be useful for understanding and predicting traffic congestion in large cities.

As a final discussion, we emphasize that all the attacks here considered are simultaneous ---the centrality measures are computed at the beginning of the attack and are not updated after each node removal. These kind of attacks corresponds to temporal scales where the node removal procedure occurs faster than the process of centrality adjustment. The extension of our analysis to updated attacks and its comparison with the present results would be an interesting research direction. 
In this line, we recall that in Erd\"os-R\'enyi networks, updating betweenness changes the universality class of the attack \cite{Almeira2020} and presents a percolation threshold close to the optimal value \cite{Morone2015}. Also, in DT networks, Norrenbrock, et al. \cite{Norrenbrock2016} showed that this attack has a percolation threshold $f_c = 0$, but the thermodynamic properties of the transition remain unknown.

\section{Conclusions}\label{sec:Conclusions}

In this article, we studied the dismantling of two-dimensional Delaunay triangulations under node removal based on betweenness centrality. We studied the breakdown of the networks in terms of percolation transitions and characterized the nature (order and criticality) of the transitions using finite-size scaling analysis. Alongside the standard definition of betweenness, we employed the so-called $\ell$-betweenness, which ignores paths longer than $\ell$ and thus, varies from local to global as the parameter $\ell$ increases.  

We found that the attack based on the full-range betweenness produces a discontinuous transition at $f_c^{\mathrm{B}}=0$. On the other hand, finite values of $\ell$ produce continuous transitions at $f_c^{\mathrm{B}\ell}>0$ that belong to the universality class of random percolation on two-dimensional lattices. By systematically varying the parameter $\ell$, we determined that the full-range behavior is recovered when $\ell$ increases with the system linear size $L$ as $\ell\sim L^{\alpha}$, with $\alpha$ close to 1, suggesting that any finite approximation of betweenness worsens the attack effectiveness by changing the transition not only quantitatively (increasing the percolation threshold) but also qualitatively (modifying its order).

\section*{Acknowledgments}

This work was partially supported by grants from CONICET (PIP 112 20150 10028), FonCyT (PICT-2017-0973), SeCyT–UNC (Argentina), and used computational resources from CCAD – Universidad Nacional de C\'ordoba (\url{http://ccad.unc.edu.ar/}), which are part of SNCAD – MinCyT, Rep\'ublica Argentina.

%\end{linenumbers}

 \bibliographystyle{elsarticle-num} 
 \bibliography{cas-refs,references}





% \end{thebibliography}
\end{document}

\begin{frontmatter}

\title{Explosive dismantling of two-dimensional random lattices under betweenness centrality attacks\\Supplementary Material}

\author[famaf, ifeg]{Nahuel Almeira}
\author[famaf, ifeg]{Juan Ignacio Perotti}
\author[ifeg]{Andrés Chacoma}
\author[famaf, ifeg]{Orlando Vito Billoni}

\affiliation[famaf]{
    organization={Facultad de Matem\'atica, Astronom\'{\i}a, F\'{\i}sica y Computaci\'on, Universidad Nacional de C\'ordoba},
    addressline={Ciudad Universitaria}, 
    city={Córdoba},
    postcode={5000}, 
    state={Córdoba},
    country={Argentina}
}
    
\affiliation[ifeg]{
    organization={Instituto de F\'{\i}sica Enrique Gaviola (IFEG-CONICET)},
    addressline={Ciudad Universitaria}, 
    city={Córdoba},
    postcode={5000}, 
    state={Córdoba},
    country={Argentina}
}

\end{frontmatter}

\section{\label{sec:SM_B} Complementary analysis for the full-range percolation transition}

\subsection{Complementary estimation for the critical exponents}

In the main text, we estimated the exponent ratio $\gamma/\nu$ from the behavior of the susceptibility $\langle s \rangle$. This ratio can be also estimated by the scaling of the fluctuations of the order parameter, defined as $\chi = \sqrt{\langle S^2\rangle - \langle S\rangle^2}$. The corresponding scaling ansatz for this metric is 

\begin{equation} \label{eq:chi_scaling}
      \chi(f, L) \sim L^{\gamma / \nu} \tilde{\chi} [(f-f_c) L^{1/\nu}].
\end{equation}

\begin{figure}[ht]
\centering
\includegraphics[scale=0.35]{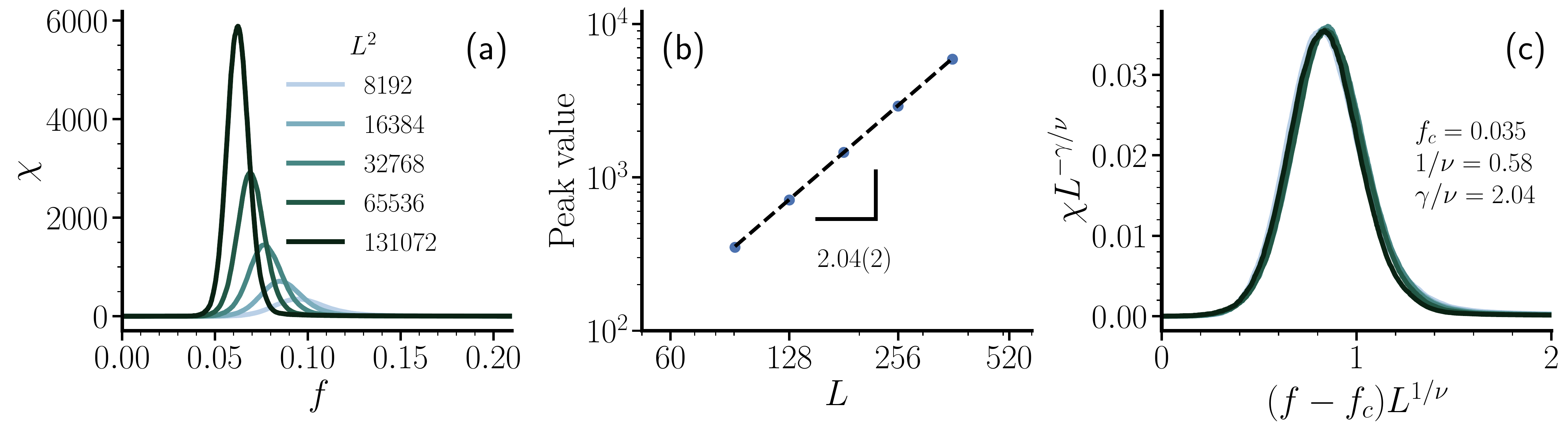}
\caption{\label{fig:collapse_chi}
\textbf{(a)} Fluctuations of the order parameter as a function of the fraction of removed nodes. \textbf{(b)} Scaling for the peaks.
\textbf{(c)} Collapse of the curves using Eq. \eqref{eq:chi_scaling}. }
\end{figure}

In Figure \ref{fig:collapse_chi} we show the curves for the fluctuations as a function of the fraction of nodes removed for different system sizes. By the scaling of the peaks,  shown in Panel \ref{fig:collapse_chi}b, we obtain $\gamma / \nu = 2.05(1)$, which is close to the value obtained from the scaling of $\langle s \rangle$, and also consistent with a first-order transition. The collapse for the curves is good, confirming Eq. \eqref{eq:chi_scaling}. 

\section{Percolation transition for $\ell$-betweenness attacks} \label{sec:Bl_percolation}

\begin{figure}
\centering
\includegraphics[scale=0.35]{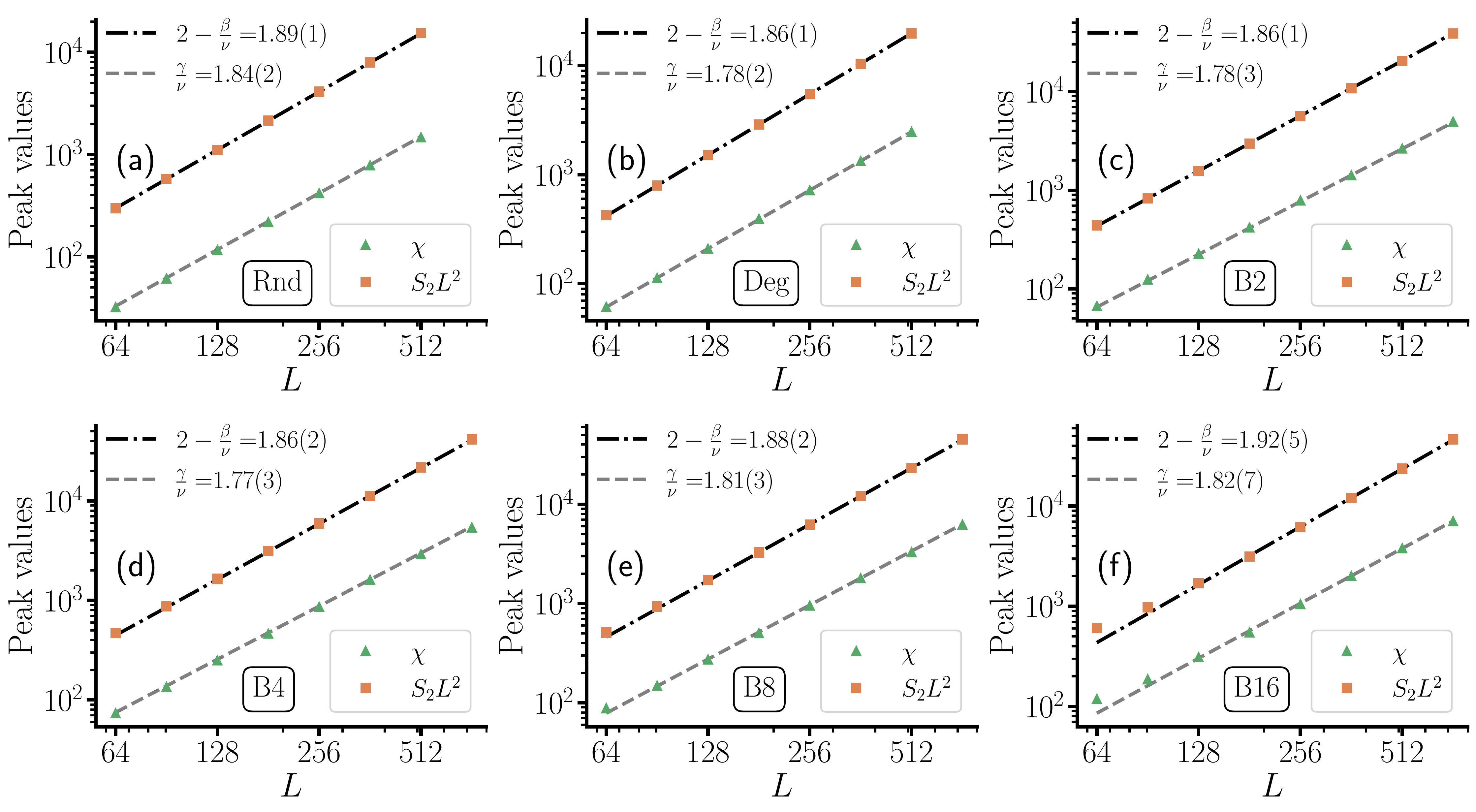}
\caption{\label{fig:beta_gamma_nu_Bl} Scaling for the peak of the second moment and second-largest cluster for $\ell$-betweenness attacks. 
For comparison, random (Rnd) node removal and initial degree attack strategy (Deg) are also included. 
As a reference, for standard percolation in 2-dimensional regular lattices, the corresponding values are $\gamma/\nu = 1.792$ and $2-\beta/\nu = 1.896$ \cite{StaufferBook}.
}
\end{figure}

In this section, we show results for the main characteristics of the percolation transitions induced by $\ell$-betweenness attacks on DT networks. In Figure \ref{fig:beta_gamma_nu_Bl} we show the scaling for the peaks of the second moment of the finite-component size distribution $M_2$ and second-largest cluster $S_2L^2$. Each peak is determined from an average of $10^3-10^4$ realizations. As it can be seen, the exponents do no change with $\ell$ and, within uncertainties, coincide with the exponents associated with standard percolation on two dimensions. Thus, the universality class of the transition does not change as long as the cutoff remains finite.

\begin{figure}
\centering
\includegraphics[scale=0.35]{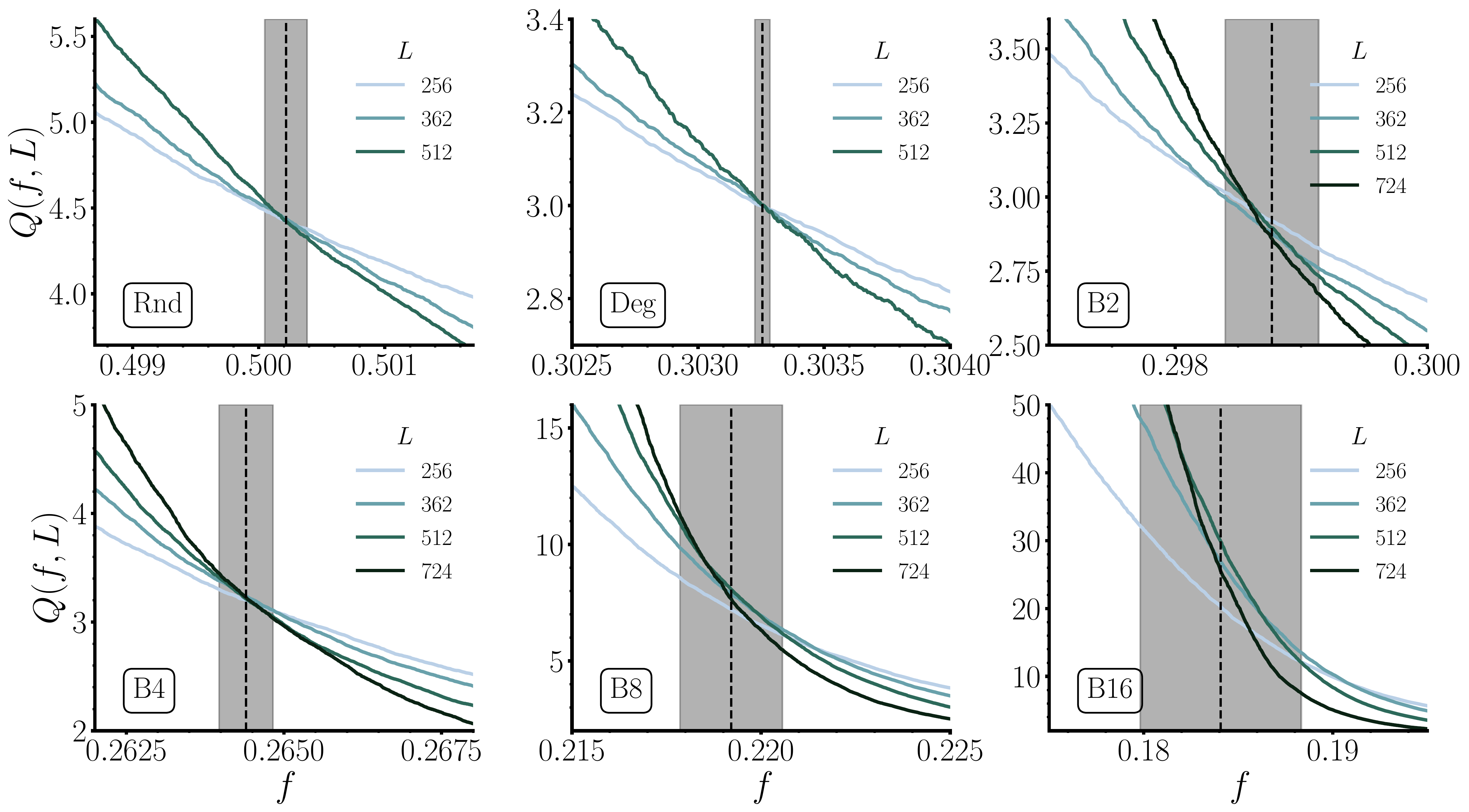}
\caption{\label{fig:crossing_Bl} Estimation of the percolation threshold for $\ell$-betweenness attacks with different cutoff lengths $\ell$, by means of the crossing method. As in Figure \ref{fig:beta_gamma_nu_Bl}, random and initial degree attack strategies are shown for comparison. Averages are taken over $10^5$ simulations and the relation $\ell < \ell^*(L)$ is satisfied for all the network sizes included in these estimation.}
\end{figure}

%In \cite{Almeira2020}, we estimated with high precision the position of the percolation threshold by using the so-called crossing method~\cite{Fan2012}. This method is based on the scaling ansatz (2), which indicates that, for a continuous percolation transition, the component ratio $S_2/S_1$ at $f=f_c$ is independent of the system size. That is, $Q(f_c, L) = S_2(f_c,L)/S_1(f_c, L) = \mathrm{constant}$. Thus, if the component ratios for different sizes are plotted versus the fraction of nodes removed, all curves should intersect at a single point corresponding to the percolation threshold.
Although the critical exponents are not sensitive to the details of the attack strategy, the percolation threshold is expected to vary in terms of the attack. As the cutoff increases, more information is obtained from the neighborhood of each node, so the percolation threshold should decrease as $\ell$ becomes larger. To check this, we computed the percolation threshold for different values of $\ell$ up to $\ell=16$. 
We employed the \emph{crossing method} as in \cite{Almeira2020}, 
which consists in computing the intersection point between the ratios $Q(f, L) = S_1(f, L)/S_2(f,L)$ of the largest and second-largest component sizes for different system sizes. In Figure \ref{fig:crossing_Bl} we show the crossings of $Q(f,L)$ for each attack. For comparison, we include random percolation (Rnd) and initial degree-based attack (Deg). In particular, for random percolation we obtain $f_c^{\mathrm{Rnd}} = 0.5002(2)$, which is consistent with the theoretical value $f_c = 1/2$ \cite{Sykes1964}. We note that, as the cutoff length increases, finite-size effects do it as well, so larger systems are needed to get a good estimation. Together with the incremental computational cost of extending the interaction radius, this makes it difficult to estimate $f_c^{\mathrm{B}\ell}$ beyond $\ell=16$.

%In Figure \ref{fig:fc_and_rc_Bl} we plot the position of each percolation threshold as a function of $\ell$. For lower cutoff values, a power-law decay of the threshold is seen, but as $\ell$ keeps increasing, the variation becomes slower. For comparison, we also include the values for the largest gap position $f_{\Delta}^{\mathrm{B}\ell}(L)$ for different network sizes. We can observe that the deviation of $f_c^{\mathrm{B}\ell}$ from a power-law matches the plateau that emerges for $f_{\Delta}^{\mathrm{B}\ell}(L)$ in large systems.

%\begin{figure}[ht]
%\centering
%\includegraphics[scale=0.4]{f_c_and_r_c_BlDT.pdf}
%\caption{\label{fig:fc_and_rc_Bl} 
%Percolation threshold $f_c^{\mathrm{B}\ell}$, and finite-size largest gap position $f_{\Delta}^{\mathrm{B}\ell}(L)$ as a function of the cutoff length $\ell$. For lower values of $\ell$ the percolation threshold decays as a power-law with exponent $\sim 0.19$. 
%As $\ell$ increases, the decay is lowered and the plateau seen for the largest gap position in large system sizes emerges.
%}
%\end{figure}

\begin{figure}[ht]
\centering
\includegraphics[scale=0.4]{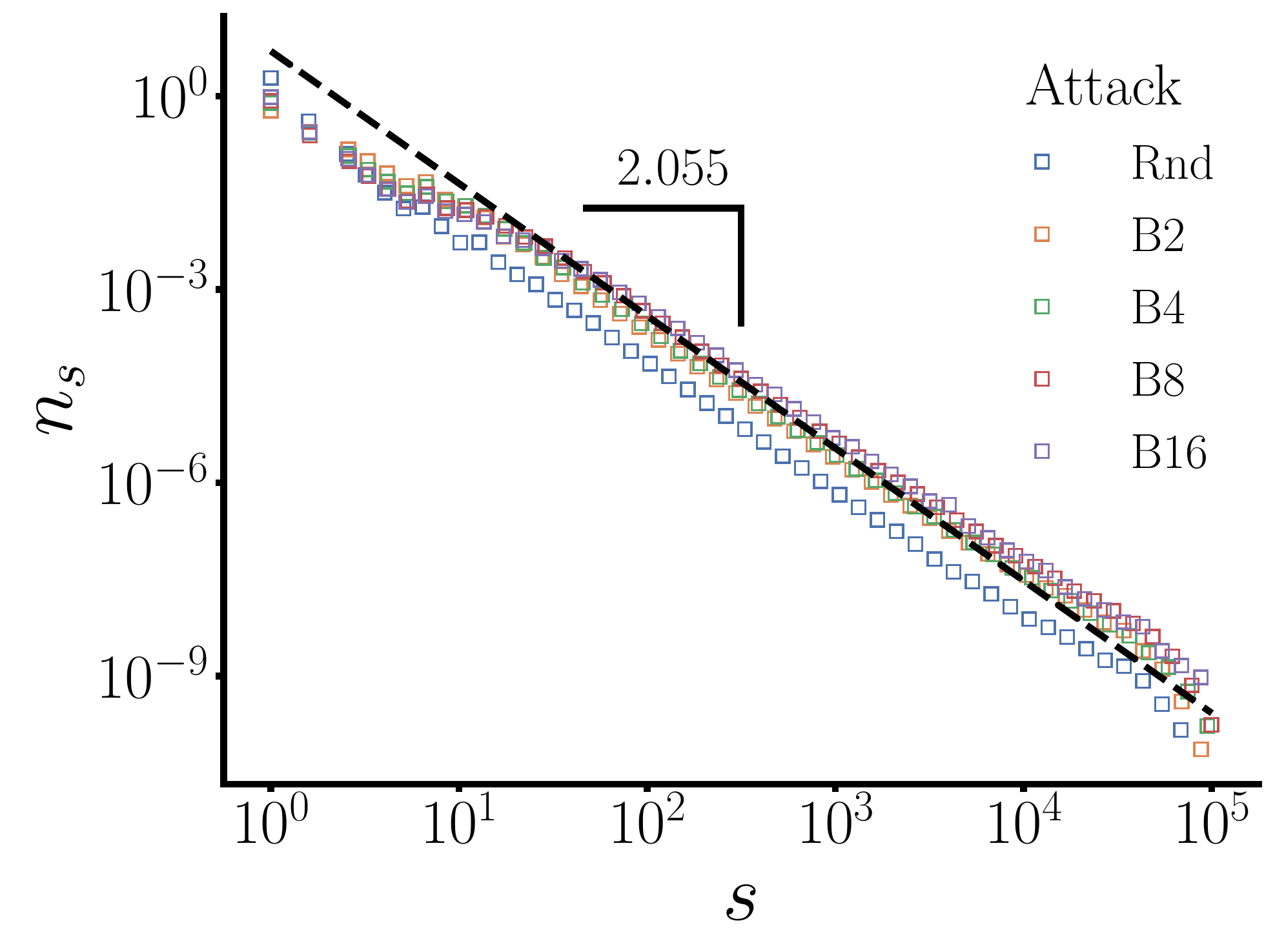}
\caption{\label{fig:Bl_ns}
Finite component size distribution for different $\ell$-betweenness attacks on networks of size $L=724$ at $f^{\mathrm{B}\ell}_c(L)$. Each histogram is generated over an average of $10^3$ networks. The dashed line represents the relation $n_s \sim s^{-\tau}$, with $\tau = 0.2055$, corresponding to random percolation on an infinite network.}
\end{figure}

To conclude, we show in Figure \ref{fig:Bl_ns} the distribution of the finite-component sizes for different cutoffs $\ell$ on large networks, where the condition $\ell\ll \ell^*(L)$ holds. We can see that in all cases the distribution is similar to random percolation, with a power-law relation $n_s\sim s^{-\tau}$, where $\tau = 2.055$ is the fisher exponent corresponding to random percolation on an infinite two-dimensional lattice.

\section{Performance comparison for range-limited betweenness algorithms}

\begin{figure}
\centering
\includegraphics[scale=0.4]{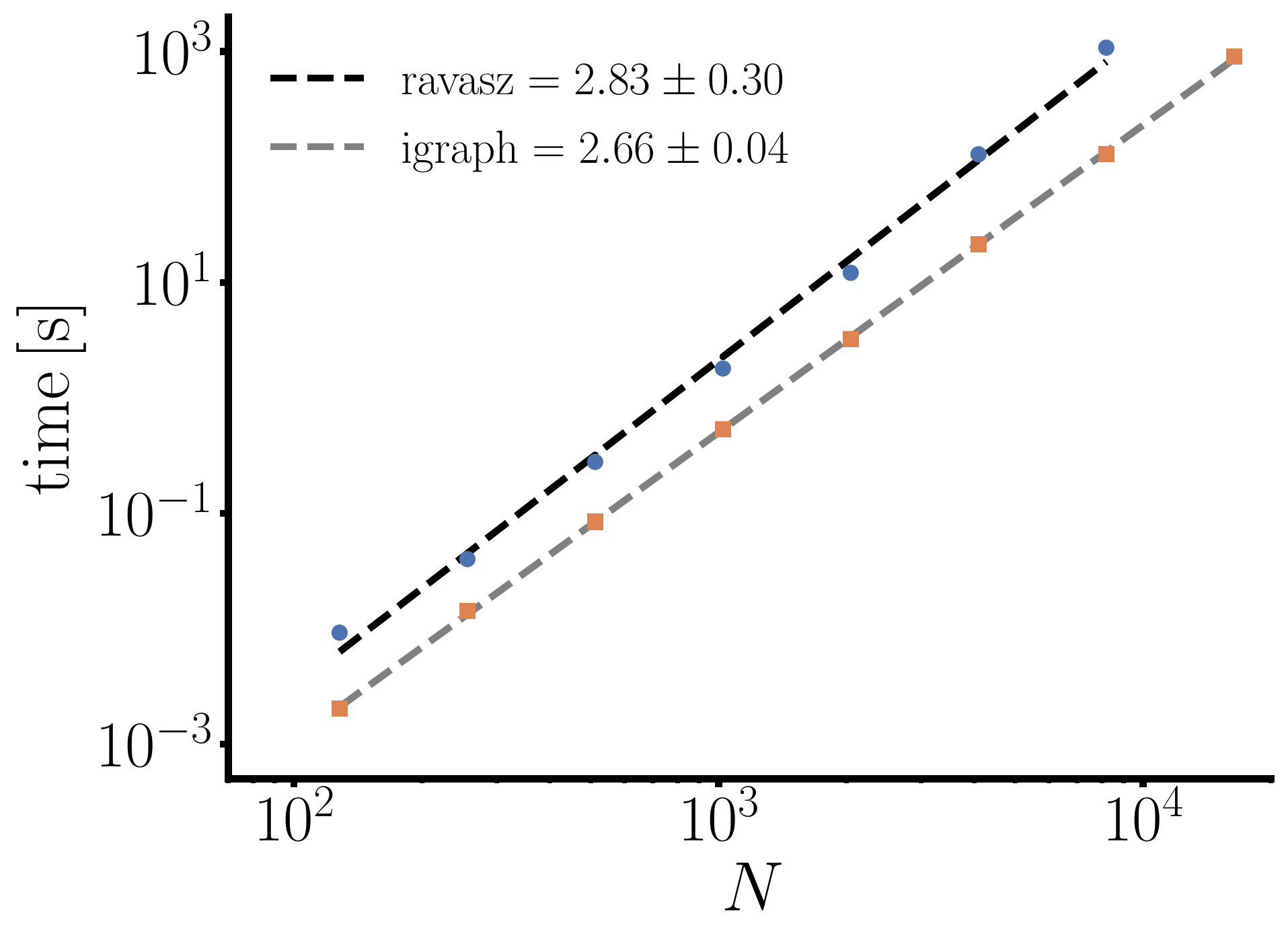}
\caption{\label{fig:time_test} Time complexity comparison of two different algorithms for computing range-limited betweenness. }
\end{figure}

As mentioned in the main text, a naive computation of betweenness has a complexity of $\mathcal{O}(N^2M)$. Brandes algorithm \cite{Brandes2001} reduces the complexity to $\mathcal{O}(NM)$, the lowest value that has been obtained so far. 
As this algorithm is based on a breadth-first search, it can be easily adapted for computing finite-range betweenness by stopping the search when the given cutoff $\ell$ is reached. 
If range-limited betweenness for multiple cutoffs is to be computed, one has to run the adapted algorithm from scratch for each value of $\ell$. Thus, the final complexity is $\mathcal{O}(\ell MN)$. In \cite{Ercsey-Ravasz2010}, the authors propose an alternative algorithm for computing range-limited betweenness, where the centrality for all cutoffs up to a given $\ell$ are computed at once. The authors show that the computational complexity $\mathcal{C}$ of their algorithm is bounded by $\mathcal{O}(NM) < \mathcal{C} <\mathcal{O}(\ell MN)$. To compare both algorithms, we performed the following experiment. For a given network size $L$, we computed all the range-limited betweenness values up to $\ell^*(L)$. We repeated the computation 10 times and averaged the computation time. The same procedure was then performed for different network sizes. As $\ell^*$ scales with the system size as $\ell^*\sim L = N^{\alpha/2}$ for DT networks, $M\sim N$, the expected worst-case complexity for both algorithms is $\mathcal{O}(N^{2+\alpha/2}) \approx \mathcal{O}(N^{2.46})$, according to the estimation of $\alpha$ given in the main text.
As we can see in Figure \ref{fig:time_test}, the numerical estimation of the complexity of both algorithms is close to the theoretical prediction. In particular, the Ercsey-Ravasz algorithm is slightly slower, indicating that the real complexity is close to the upper bound, as the authors discuss in \cite{Ercsey-Ravasz2010}. 
Our performance test thus indicates that there is no gain in computation time for using the Ercsey-Ravasz algorithm. Actually, using the standard Brandes algorithm has another advantage, which is that we can omit certain values of $\ell$ when estimating $\ell^*$ and save some computation time.

 \bibliographystyle{elsarticle-num} 
 \bibliography{cas-refs,references}





% \end{thebibliography}